\documentclass[12pt]{article}
\pdfoutput=1

\usepackage{amsmath,amssymb,amsfonts,graphicx,amsfonts}
\usepackage{color}
\usepackage[hidelinks]{hyperref}
\usepackage{ascmac}

\allowdisplaybreaks[4]

\date{}
\begin{document}
\begin{flushright}
DMUS-MP-21/02
\end{flushright}

\vspace{0.1cm}

\begin{center}
  {\large
Bulk geometry in gauge/gravity duality and color degrees of freedom\\
  }
\end{center}
\vspace{0.1cm}
\vspace{0.1cm}
\begin{center}

Masanori Hanada

\vspace{0.3cm}

Department of Mathematics, University of Surrey, Guildford, Surrey, GU2 7XH, UK\\

\end{center}

\vspace{1.5cm}

\begin{center}
  {\bf Abstract}
\end{center}

U($N$) supersymmetric Yang-Mills theory naturally appears as the low-energy effective theory of a system of $N$ D-branes and open strings between them.
Transverse spatial directions emerge from scalar fields, which are $N\times N$ matrices with color indices;
roughly speaking, the eigenvalues are the locations of D-branes.  
In the past, it was argued that this simple `emergent space' picture cannot be used in the context of gauge/gravity duality, 
because the ground-state wave function delocalizes at large $N$, leading to a conflict with the locality in the bulk geometry. 
In this paper we show that this conventional wisdom is not correct: the ground-state wave function does not delocalize, and there is no conflict with the locality of the bulk geometry. 
This conclusion is obtained by clarifying the meaning of the `diagonalization of a matrix' in Yang-Mills theory, which is not as obvious as one might think.   
This observation opens up the prospect of characterizing the bulk geometry via the color degrees of freedom in Yang-Mills theory, 
all the way down to the center of the bulk.    

\newpage
\tableofcontents
\section{Introduction}
\hspace{0.51cm}
The low-energy effective dynamics of $N$ D$p$-branes parallel to each other and open strings connecting them can be described by $(p+1)$-dimensional maximally supersymmetric Yang-Mills (SYM) theory with the U($N$) gauge group~\cite{Witten:1995im}. SYM theory has $9-p$ scalar fields $X_I$ ($I=1,2,\cdots,9-p$) which are $N\times N$ Hermitian matrices.
When all scalar fields can be (almost) simultaneously diagonalized, diagonal elements of the matrices are interpreted as the positions of D$p$-branes in the transverse directions ($(X_1^{ii},\cdots,X_{9-p}^{ii})\in\mathbb{R}^{9-p}$ is regarded as the coordinate of the $i$-th D-brane) and the off-diagonal elements $X_I^{ij}$ are interpreted as the amount of the open-string excitations connecting $i$-th and $j$-th D$p$-branes. If the matrices are not (almost) simultaneously diagonalizable but can be taken block diagonal, each block is a bound state of D-branes and strings. 

Matrix Theory conjecture~\cite{Banks:1996vh} claims the $(0+1)$-dimensional SYM --- D0-brane quantum mechanics --- is not just a low-energy effective theory, but it actually describes M-theory in certain parameter region. Bound states, or equivalently non-commutative blocks, are interpreted as various objects such as graviton, higher-dimensional D-brane and black hole. 
If we separate the $(N,N)$ component from others and see how they interact, we can study the geometry formed by $(N-1)$ D-branes and strings between them, by using the $N$-th D-brane as a probe. 

\subsubsection*{A puzzle}
\hspace{0.51cm}
Can the same geometric picture be valid in the Maldacena-type gauge/gravity duality~\cite{Maldacena:1997re,Itzhaki:1998dd}? 
Naively, we would expect that this simple mechanism of the emergent space works as follows. In $(3+1)$-dimensional super Yang-Mills theory, there are six scalars with which the coordinate in $\mathbb{R}^6$ can be specified. This, and $\mathbb{R}^{1,3}$ along which D-branes are extended, give ten-dimensional spacetime. D-branes and open strings can interact with each other and nontrivial metric can be induced effectively. $\mathbb{R}^{1,3}$  and the radial coordinate of $\mathbb{R}^6$ form AdS$_5$, and the angular part of $\mathbb{R}^6$ is S$^5$.
We can imagine similar stories for SYM in different dimensions.  
It would be nice if such a simple mechanism can actually work. However it is widely believed that this picture does not work, or at least a more sophisticated approach is required; see e.g.,~\cite{Susskind:1998vk,Polchinski:1999br,Heemskerk:2012mn,Maldacena:2018vsr,Iizuka:2001cw}.
Such skepticism is based on the observation that the matrices are highly non-commutative in the region where weakly-coupled string theory is valid, and the notion of `location' is not apparent there. This can also be phrased that the bound state of D-branes and matrices are very big compared to the counterpart in the gravity side. Later in this paper, we will show that this argument is not correct and the `location' can actually make sense. But for now let us follow the reasoning in the previous references. 
Let us consider the D0-brane matrix model\footnote{We impose the traceless condition for each matrix so that the bound state is centered around the origin of $\mathbb{R}^9$.
} with the following normalization:
\begin{align}
L
=
{\rm Tr}\left(
\frac{1}{2}(D_tX_I)^2
+
\frac{g^2}{4}[X_I,X_J]^2
+
{\rm fermion\ part}
\label{action:BFSS}
\right). 
\end{align}
Here $D_tX_I=\partial_tX_I-i[A_t,X_I]$ is the gauge-covariant derivative.  
In the 't Hooft large-$N$ limit ('t Hooft coupling $\lambda=g^2N\sim N^0$, energy $E\sim N^2$) and at sufficiently strong coupling ($\lambda\gg (E/N^2)^3$), type IIA supergravity is a good dual description~\cite{Itzhaki:1998dd}. 
In the 't Hooft large-$N$ limit, the expectation value 
$\langle {\rm Tr}X^2_I\rangle$ is of order $N^2$, at any temperature including zero and any coupling. 
At zero temperature, all contributions are from zero-point fluctuations.  
If we diagonalize each $X_I$, the eigenvalues are of order $\sqrt{N}$. So the bound state is parametrically large at large $N$. 
If we take the 't Hooft coupling $\lambda=g^2N$ to be large, then the eigenvalues scale as $\lambda^{-1/6}\sqrt{N}$ at sufficiently low temperature ($T\ll\lambda^{1/3}$).
This is larger than the size of the black hole (black zero-brane) sitting at the center of the bulk geometry, and completely covers the region where weakly-coupled string theory is valid. 
When $X_{I=1}$ is diagonalized, $X_{I=2,3,\cdots,9}$ are not diagonal at all, and the off-diagonal elements dominate $\langle {\rm Tr}X^2_{I=2,\cdots,9}\rangle$. 
Hence the `location of D-brane' is not a crisp notion when weakly-coupled string theory is valid. 
The same argument holds in any gauge theory in the 't Hooft limit, including $(3+1)$-dimensional SYM; 
when $O(N^0)$ is expected from the gravity picture, $O(\sqrt{N})$ is obtained in the gauge theory side. 
This has been regarded as an obstruction for the sub-AdS-scale bulk reconstruction in AdS/CFT correspondence. 
\subsubsection*{A resolution}
\hspace{0.51cm}
In this paper, we show that the size of the bound state in the gauge theory side is actually much smaller, and the `location' can make sense. 
Whether the metric expected in the holographic duality actually emerges is a separate issue, which will not be discussed in this paper. 
(We will suggest a few future directions aiming for the verification of the emergence of the local bulk geometry.)

The starting point of our discussion is this question: 
\begin{center}
\textit{What do we mean by the ``diagonalization of a matrix"?}
\end{center}
Of course, when an $N\times N$ Hermitian matrix is given, there is no ambiguity; it is a well-defined linear-algebra problem. 
However, because we are talking about a physics problem, we have to make sure what is the `matrix' suitable for the problem under consideration. 
Namely, we have to answer the following question: 
\begin{center}
\textit{What is the ``matrix" that characterizes the bulk geometry?}
\end{center}

The argument in the past implicitly used one of the following two pictures:
(1) interpret a typical configuration ($\sim$ master field) in the path integral as a `bound state', 
or
(2) interpret a typical result of the measurement of $X_{I,ij}$ (which is a coordinate eigenstate described by the coordinates $X_{I,ij}$, i.e.,  a state $|X\rangle$ that satisfies $\hat{X}_{I,ij}|X\rangle=X_{I,ij}|X\rangle$ for all $I,i,j$) as a `bound state'.\footnote{Strictly speaking, we have to take into account the fermions as well.}
Either way, there are c-number matrices $X_{I,ij}$, so we can `diagonalize' one of them and define `eigenvalues of a matrix $X_{I,ij}$'. 
In the picture (2), $X_{I,ij}$ is the `eigenvalue of operator $\hat{X}_{I,ij}$'. 
Both (1) and (2) fail in more or less the same manner, so let us focus on (2) below for concreteness. 
Furthermore, we consider the D0-brane theory that has nine scalars $X_{I=1,2,\cdots,9}$. (The generalizations to other theories are straightforward.)

By assumption, we are interested in low-energy states including the ground state. 
A coordinate eigenstate cannot be a low-energy state due to the uncertainty principle; instead, we must consider linear combinations of coordinate eigenstates, such as a wave packet. 
In general, a low-energy state $|\Phi\rangle$ is written in terms of the coordinate eigenstates $|X\rangle$ as
\begin{align}
|\Phi\rangle
=
\int_{\mathbb{R}^{9N^2}} 
dX |X\rangle\langle X|\Phi\rangle
\equiv
\int_{\mathbb{R}^{9N^2}} 
dX\Phi(X) |X\rangle.  
\end{align}
The wave function $\Phi(X)$ has to be extended smoothly in $\mathbb{R}^{9N^2}$ to some extent.\footnote{
We emphasize that the wave packet under consideration is in $\mathbb{R}^{9N^2}$ and \textit{not} in $\mathbb{R}^{9}$.
The bound state of D-branes and open strings can be extended in $\mathbb{R}^{9}$, but it is a completely different story. 
In the past, these two completely different notions ---- 
a wave packet extended in $\mathbb{R}^{9N^2}$, 
and a bound state extended in $\mathbb{R}^{9}$ --- 
were not properly distinguished.
}
Hence `the eigenvalue of operator $\hat{X}_{I,ij}$' is not well-defined, 
and a naive `diagonalization' based on the intuition from coordinate eigenstates is not well-defined either.
That $\langle {\rm Tr}X^2_I\rangle$ is of order $N^2$ does not necessarily mean
`the eigenvalues of $X_I$' are of order $\sqrt{N}$; 
actually \textit{the very notion of the `eigenvalues' has to be defined more carefully}.
In order to define the `coordinate of D-branes' in $\mathbb{R}^9$,
we have to define the `coordinate of matrices' in $\mathbb{R}^{9N^2}$.

In fact there is a very standard way to introduce the `coordinate of matrices': 
if $\Phi(X)$ is a wave packet around $Y_{I,ij}$, 
the center of the wave packet $Y_{I,ij}$ is a natural `coordinate of matrices'.\footnote{
This way of introducing a `coordinate of matrices' does not work for more generic states, such as a superposition of multiple wave packets. 
This is not a bug, this is a generic feature of quantum mechanics. 
} 
This point is explained in Sec.~\ref{sec:MM-Hamiltonian}.
We will see that this $Y_{I,ij}$ is naturally related to the locations of D-branes and open-string excitations. 
It turns out that the ground state is a wave packet localized around the origin of $\mathbb{R}^{9N^2}$, i.e., $Y_{I,ij}=0$. 
Along each direction of $\mathbb{R}^{9N^2}$, the width of the wave packet is of order $N^0$. 
This is the reason that $\langle {\rm Tr}X^2_I\rangle$ is of order $N^2$. 
The ground state is gauge-invariant, i.e., it is impossible to change the shape of the wave packet via gauge transformation. 
It is perfectly consistent with a simple and natural interpretation: in the ground state, all D0-branes are sitting at the origin of the bulk, and no open string is excited.   
\subsubsection*{The organization of this paper}
\hspace{0.51cm}
This paper is organized as follows.
In Sec.~\ref{sec:MM-Hamiltonian}, we consider matrix models. 
To make the logic transparent, we use the canonical quantization and quantum states, rather than the path-integral formalism. 
All the essence which can readily be generalized to Yang-Mills theory in any dimension can be understood just by considering the Gaussian matrix model, which is the subject of Sec.~\ref{sec:Gaussian}. 
We show that the wave function does not delocalize, and probes can be introduced in a very standard manner. 
In Sec.~\ref{sec:MM-finite-coupling}, we will see how simple results obtained for the Gaussian matrix model are generalized to interacting theories. 
In Sec.~\ref{sec:D0-MM}, we consider the D0-brane matrix model and dual gravity description. 
Sec.~\ref{sec:D0-MM} is rather speculative, because we have not yet fully understood the dynamics of the model. 
In Sec.~\ref{sec:4dSYM}, we consider $(3+1)$-dimensional super Yang-Mills and AdS$_5$/CFT$_4$ correspondence. 
Potentially interesting future directions are discussed in Sec.~\ref{sec:future-directions}. 
\section{Matrix Model via canonical quantization}\label{sec:MM-Hamiltonian}
\hspace{0.51cm}
In this section, we consider the matrix model. Before studying the D0-brane matrix model, let us consider 
a simpler example, a nine-matrix model with the following Lagrangian (with the Minkowski signature):
\begin{align}
L
=
{\rm Tr}\left(
\frac{1}{2}(D_tX_I)^2
-
\frac{1}{2}X_I^2
+
\frac{g^2}{4}[X_I,X_J]^2
\right). 
\label{Lagrangian-simple-MM}
\end{align}
The zero-coupling limit (the Gaussian matrix model) is analytically solvable, and we can understand everything explicitly. 
At strong coupling, the quadratic term $-\frac{1}{2}X_I^2$ is negligible and this model reduces to the bosonic part of the D0-brane matrix model. 
This model was studied in detail via lattice Monte Carlo simulation~\cite{Bergner:2019rca,Watanabe:2020ufk}. 
While this theory does not have a weakly-coupled gravity dual, all the essential points can be illuminated by using this example, without having technical complications. We will discuss the D0-brane matrix model toward the end of this section. 

From the Lagrangian \eqref{Lagrangian-simple-MM}, we obtain the Hamiltonian
\begin{align}
\hat{H}
=
{\rm Tr}\left(
\frac{1}{2}\hat{P}_I^2
+
\frac{1}{2}\hat{X}_I^2
-
\frac{g^2}{4}[\hat{X}_I,\hat{X}_J]^2
\right). 
\end{align}
Because we are studying the {\it gauged} matrix model, the physical states are gauge-invariant. 
Let us denote the Hilbert space of gauge-singlet states as ${\cal H}_{\rm inv}$.  
When the matrices are related to D-branes and strings, our brains tend to think in the `Higgsed' picture, namely we often consider the situation that diagonal elements are large and well-separated. This intuition uses non-singlet states. 
Hence let us also consider the extended Hilbert space ${\cal H}_{\rm ext}$ that contains non-singlet states as well. 
The partition function associated with the canonical ensemble at temperature $T$ can be written as
$Z(T)={\rm Tr}_{{\cal H}_{\rm inv}}e^{-\hat{H}/T}$, where ${\rm Tr}_{{\cal H}_{\rm inv}}$ is the trace over gauge singlets. 
We can also write it by using the trace in the extended Hilbert space as  
$Z(T)=\frac{1}{{\rm vol U}(N)}\int dU{\rm Tr}_{{\cal H}_{\rm ext}}(\hat{U}e^{-\hat{H}/T})$. Here $U$ is an element of U($N$), and the integral is taken by using the Haar measure. The operator $\hat{U}$ enforces the gauge transformation corresponding to a group element $U$, and  
$\int dU\hat{U}$ serves as the projector to ${\cal H}_{\rm inv}$.
In terms of ${\cal H}_{\rm ext}$, `gauge fixing' can naturally be understood as in the path integral formalism.  
See Appendix~\ref{app:Hinv-vs-Hext} for more details. 

Each state $|\Phi\rangle$ can be expressed by using the wave function in the coordinate basis, 
\begin{align}
\langle X|\Phi\rangle
=
\Phi(X), 
\end{align}
where $\Phi(X)$ is a function of $9N^2$ variables $X_I^{ij}$. 
If $\Phi(X)$ is well-localized wave packet in the $9N^2$-dimensional space centered around $X_I^{ij}= x_{I,i}\delta_{ij}$, 
then $\vec{x}_i=(x_{1,i},\cdots,x_{9,i})\in\mathbb{R}^9$ is naturally interpreted as `the location of the $i$-th D-brane'.

Let us use the generators of U($N$), which are denoted by $\tau_\alpha$ and normalized as ${\rm Tr}(\tau_\alpha\tau_\beta)=\delta_{\alpha\beta}$. We can write $\hat{X}_I$ and $\hat{P}_I$ as 
$\hat{X}_I^{ij}=\sum_\alpha\hat{X}_I^\alpha\tau_\alpha^{ij}$
and
$\hat{P}_I^{ij}=\sum_\alpha\hat{P}_I^\alpha\tau_\alpha^{ij}$. 
The commutation relation is 
\begin{align}
[\hat{X}_{I}^\alpha,\hat{P}_J^\beta]=i\delta_{IJ}\delta^{\alpha\beta}, 
\qquad
[\hat{X}_{I}^\alpha,\hat{X}_J^\beta]
=
[\hat{P}_{I}^\alpha,\hat{P}_J^\beta]
=
0.
\end{align}
By using the annihilation operators $\hat{a}_{I,\alpha}=\frac{\hat{X}_{I,\alpha}-i\hat{P}_{I,\alpha}}{\sqrt{2}}$
and creation operators $\hat{a}_{I,\alpha}^\dagger$, we can construct the Fock basis for ${\cal H}_{\rm ext}$.  
For each $(I,\alpha)$, the number operator is defined by $\hat{n}_{I,\alpha}=\hat{a}^\dagger_{I,\alpha}\hat{a}_{I,\alpha}$, 
and the Fock state is defined as the eigenstate of the number operator, $\hat{n}_{I,\alpha}|n\rangle_{I,\alpha}=n_{I,\alpha}|n\rangle_{I,\alpha}$.
Specifically, the Fock vacuum $|0\rangle_{I,\alpha}$ is specified by $\hat{a}_{I,\alpha}|0\rangle_{I,\alpha}=0$, 
and the excited states are constructed as $|n\rangle_{I,\alpha}=\frac{(\hat{a}_{I,\alpha}^\dagger)^n}{\sqrt{n!}}|0\rangle_{I,\alpha}$.  
Then we can simply take the tensor product, 
\begin{align}
|\{n_{I,\alpha}\}\rangle
=
\otimes_{I,\alpha}|n_{I,\alpha}\rangle_{I,\alpha},  
\end{align}
to obtain the orthonormal basis of ${\cal H}_{\rm ext}$. 
If we take a specific set of $\{n_{I,\alpha}\}$ in which the diagonal elements are highly excited while the off-diagonal elements are not, then such state is analogous to the `Higgsed' states. Indeed, by taking a linear combination of such states, we can build a wave packet localized about $X_{I,ii}\neq 0$ ($i=1,2,\cdots,N$) and $X_{I,ij}=0$ ($i\neq j$).  

The U($N$) transformation is defined by 
\begin{align}
\hat{X}_{I,ij}\longrightarrow (U\hat{X}_IU^{-1})_{ij} = \sum_{k,l=1}^NU_{ik}\hat{X}_{I,kl}U^{-1}_{lj}
\end{align}
and 
\begin{align}
\hat{P}_{I,ij}\longrightarrow (U\hat{P}_IU^{-1})_{ij} = \sum_{k,l=1}^NU_{ik}\hat{P}_{I,kl}U^{-1}_{lj}. 
\end{align}
Creation and annihilation operators are transformed in the same manner.   
With the adjoint index $\alpha$, the transformation rule is 
\begin{align}
\hat{X}_\alpha
=
{\rm Tr}(\hat{X}\tau_\alpha)
&\longrightarrow
\hat{X}^{(U)}_\alpha = {\rm Tr}((U\hat{X} U^{-1})\tau_\alpha), 
\nonumber\\
\hat{P}_\alpha
=
{\rm Tr}(\hat{P}\tau_\alpha)
&\longrightarrow
\hat{P}^{(U)}_\alpha = {\rm Tr}((U\hat{P} U^{-1})\tau_\alpha),   
\nonumber\\
\hat{a}_\alpha
=
{\rm Tr}(\hat{a}\tau_\alpha)
&\longrightarrow
\hat{a}^{(U)}_\alpha = {\rm Tr}((U\hat{a} U^{-1})\tau_\alpha).    
\label{eq:gauge-transf-operators}
\end{align} 
The Fock vacuum $|\{0\}\rangle=\otimes_{I,\alpha}|0\rangle_{I,\alpha}$ is U($N$)-invariant, and the excited states transform as 
\begin{align}
|\{n_{I,\alpha}\}\rangle
=
\left(\prod_{I,\alpha}\frac{(\hat{a}_{I,\alpha}^\dagger)^{n_{I,\alpha}}}{\sqrt{n_{I,\alpha}!}}\right)|\{0\}\rangle
\longrightarrow
\left(\prod_{I,\alpha}\frac{(\hat{a}_{I,\alpha}^{(U)\dagger})^{n_{I,\alpha}}}{\sqrt{n_{I,\alpha}!}}\right)|\{0\}\rangle. 
\label{gauge-transf-Fock-state}
\end{align}
From each non-singlet state, a U($N$)-invariant state is obtained by averaging over all group elements of U$(N)$
and then properly normalizing the norm. 
\subsection{Gaussian matrix model}\label{sec:Gaussian}
\hspace{0.51cm}
Let us consider the case of $g^2=0$, i.e., the Gaussian matrix model. This example is solvable, and contains all the essence.
\subsubsection{Ground state (completely-confined state)}
\hspace{0.51cm}
In the free limit ($g^2=0$), the Hamiltonian is 
\begin{align}
\hat{H}_{\rm Gaussian}
=
\sum_{\alpha=1}^{N^2}\left(
\frac{1}{2}\hat{P}_{I,\alpha}^2
+
\frac{1}{2}\hat{X}_{I,\alpha}^2
\right). 
\end{align} 
The ground state is the Fock vacuum:
\begin{align}
|{\rm ground\ state}\rangle
=
|\{0\}\rangle
=
\otimes_{I,\alpha}|0\rangle_{I,\alpha}. 
\end{align}
The vacuum expectation value of $\sum_I{\rm Tr}\hat{X}_I^2$ is $\frac{9}{2}N^2$ due to the zero-point fluctuation. 
Hence based on the conventional wisdom one would conclude that the size of the ground-state wave function is of order $\sqrt{N}$. 
However this is actually not the case. 
Because the Fock vacuum of each harmonic oscillator is represented by the Gaussian wave function $\langle X_{I,\alpha}|0\rangle_{I,\alpha}=\frac{e^{-X_{I,\alpha}^2/2}}{\pi^{1/4}}$, the wave function describing all matrix entries is
\begin{align}
\langle X|{\rm ground\ state}\rangle
&=
\frac{1}{\pi^{9N^2/4}}\exp\left(-\frac{1}{2}\sum_{I,\alpha}X_{I,\alpha}^2\right)
\nonumber\\
&=
\frac{1}{\pi^{9N^2/4}}\exp\left(-\frac{1}{2}\sum_{I}{\rm Tr}X_{I}^2\right).  
\end{align}
This is manifestly U($N$)-invariant. 
The size of the wave function is the same for all matrix entries. 
We cannot arrange the ground-state wave function such that we can observe a large value of a diagonal element (more specifically, of order $\sqrt{N}$) with large probability, in any `gauge'.  
Typically ${\rm Tr}X_I^2$ is of order $N^2$, but this is because all the entries can take order $N^0$ values, and the probability of at least one eigenvalue becoming of order $\sqrt{N}$ scales roughly as $e^{-N}$, which is negligible at large $N$.
This state is a well-localized wave packet in the $9N^2$-dimensional space centered around $X_I^{ij}=0$.~\footnote{
It may be instructive to rephrase it as follows.
Imagine a uniform probability distribution on a sphere with radius $R$ in $D$ dimensions parametrized by $x_1,\cdots,x_D$.
By integrating out $x_2,\cdots,x_D$, we obtain the distribution of $x_1$ scaling as
$\rho(x_1)\sim\left(1-\frac{x_1^2}{R^2}\right)^{(D-2)/2}$.
In the matrix model, we roughly have a situation that $D\sim N^2\to\infty$, $R\sim N\to\infty$, which leads to $\rho(x_1)\sim e^{-\frac{Dx_1^2}{2R^2}}\sim e^{-x_1^2}$.
Therefore, if the radius increases with dimensions as $D\sim R^2$, large radius does not mean the delocalization. 
}  
Namely, all D-branes are sitting at the origin, and no open string is excited.   
Note that the full U($N$)-invariance is a natural property of $N$-coincident D-branes without open string excitations~\cite{Witten:1995im}. 

Of course, each $|X\rangle$ is not U($N$)-invariant, and we can `choose a gauge' e.g., in which $X_1$ is diagonal, if we like. 
However, the linear combination $\int dX |X\rangle\langle X|{\rm ground\ state}\rangle$ is U($N$)-invariant
and there is no way to choose any `gauge'.
As far as we consider low-energy states, it is meaningless to talk about the eigenvalue distribution of the coordinate eigenstate $|X\rangle$.  

It may be instructive to emphasize the difference between two kinds of the `size of bound state' that were not properly distinguished in the past. 
The first one is the distribution of D-branes (diagonal elements) that can be read off from the center of the wave packet. 
This is defined in $\mathbb{R}^{9}$.
The second one is the width of the wave packet in $\mathbb{R}^{9N^2}$.   
These two notions correspond to the `slow modes' and `fast modes' in the references, respectively. 
We have seen that, for the ground state, the latter is of order $N^0$. All D-branes are sitting at the origin, so the former is zero.
\subsubsection{Coherent states}\label{sec:coherent-state}
\hspace{0.51cm}
Perhaps it is not easy to grasp the essence of the statement just by looking at the ground state. 
Let us examine the coherent states, which nicely illuminate the important points. 

We can put the wave packet at any point in $\{Y_{I,\alpha}\}\in\mathbb{R}^{9N^2}$, just by acting the translation operator:
\begin{align}
|{\rm wave\ packet\ at\ } \{Y_{I,\alpha}\}\rangle
&=
e^{-i\sum_{I=1}^9\sum_{\alpha=1}^{N^2}Y_{I,\alpha}\hat{P}_{I,\alpha}}|{\rm ground\ state}\rangle
\nonumber\\
&=
e^{-i\sum_{I=1}^9{\rm Tr}(Y_I\hat{P}_I)}|{\rm ground\ state}\rangle. 
\label{eq:wave-packet-generic}
\end{align} 
A more generic wave packet with nonzero momentum is 
\begin{align} 
e^{-i\sum_{I=1}^9{\rm Tr}(Y_I\hat{P}_I-Q_I\hat{X}_I)}|{\rm ground\ state}\rangle.
\label{eq:wave-packet-generic-with-momentum}
\end{align}  
Below we mainly focus on the case of $Q_I=0$ for simplicity. 

\begin{figure}[htbp]
  \begin{center}
   \includegraphics[width=100mm]{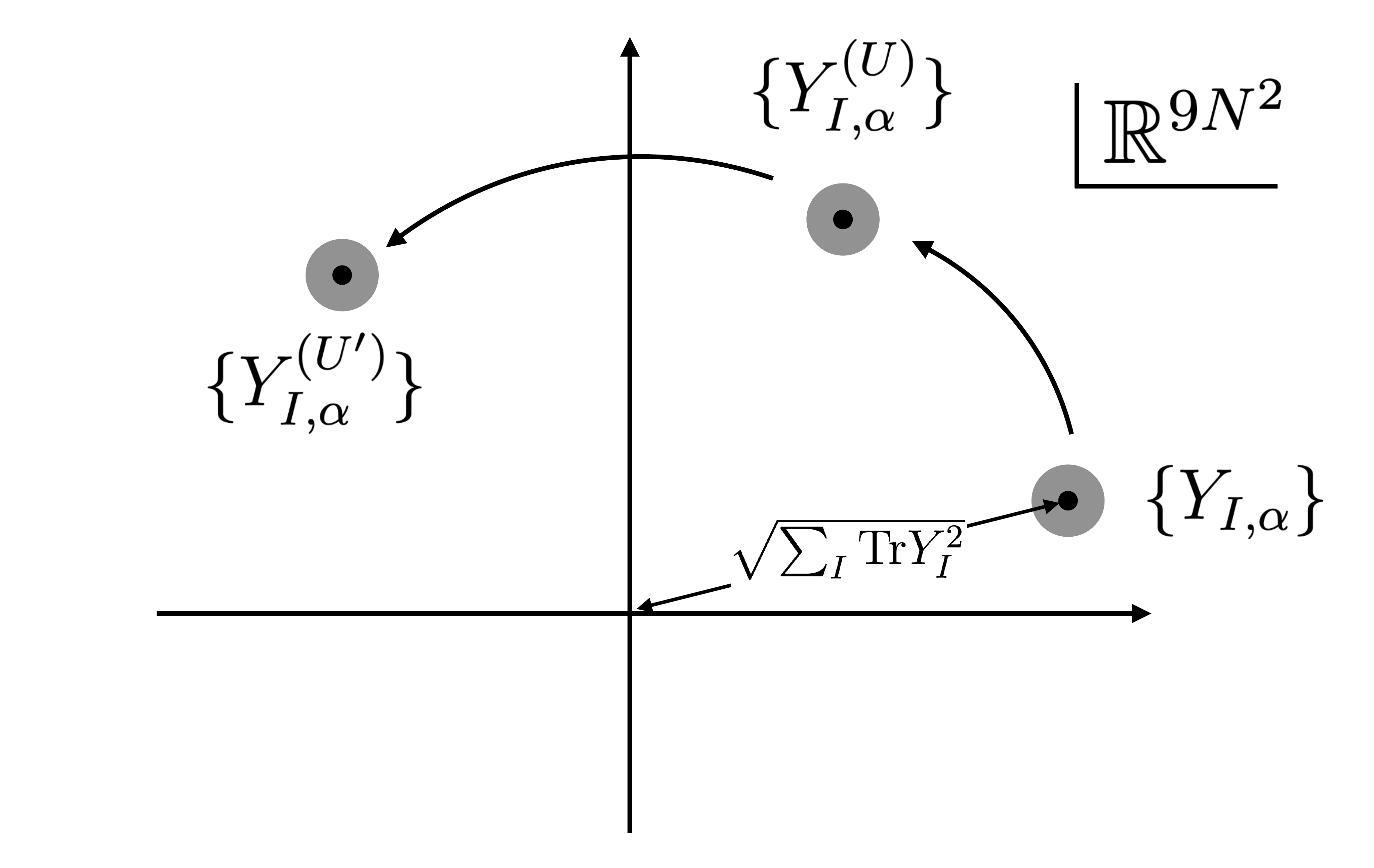}
  \end{center}
  \caption{
  The coherent state in $\mathbb{R}^{9N^2}$ defined by \eqref{eq:wave-packet-generic}.
  Each gray disk and black point represent a wave packet and its center, respectively. 
  Under the gauge transformation, the location of the center moves, but the shape of the wave packet in $\mathbb{R}^{9N^2}$ does not change. 
      }\label{fig:wave-packet}
\end{figure}

The center of the wave packet $\{Y_{I,\alpha}\}\in\mathbb{R}^{9N^2}$
describes the D-brane configuration, which corresponds to the `slow mode' in the references. 
It can change via the U($N$) transformation as 
\begin{align}
|{\rm wave\ packet\ at\ } \{Y_{I,\alpha}\}\rangle
&\longrightarrow
|{\rm wave\ packet\ at\ } \{Y^{(U)}_{I,\alpha}\}\rangle, 
\end{align} 
where
\begin{align}
Y^{(U)}_{I,ij}
=(U^{-1}Y_IU)_{ij}. 
\end{align} 
See Fig.~\ref{fig:wave-packet} for a visual sketch. 
Therefore it makes sense to talk about the diagonalization of the slow mode $Y$.~\footnote{ 
The eigenvalues of $Y_{I,ij}$ are gauge-invariant, 
and the distance from the origin in $\mathbb{R}^{9N^2}$, that can be expressed as $\sqrt{\sum_I{\rm Tr}Y_I^2}$, is also gauge-invariant.} 
However the width of the wave packet, that comes from $|{\rm ground\ state}\rangle$,
does not change via the U($N$) transformation; 
see Fig.~\ref{fig:wave-packet} again.
This is because 
\begin{align}
&
\langle {\rm wave\ packet\ at\ } \{Y_{I,\alpha}\}|
(\hat{X}_{I,\alpha}-Y_{I,\alpha})^k
|{\rm wave\ packet\ at\ } \{Y_{I,\alpha}\}\rangle
\nonumber\\
&\hspace{1cm}=
\langle {\rm ground\ state}|
\hat{X}_{I,\alpha}^k
|{\rm ground\ state}\rangle
\end{align}
holds for each $(I,\alpha)$ and any $k$, and the right hand side is gauge-invariant. 
This part picks up the quantum fluctuation, which corresponds to the `fast mode' in the references; 
hence it does not make sense to talk about the diagonalization of the fast mode. 

If we measure the coordinate in $\mathbb{R}^{9N^2}$, 
we get a localized distribution around $\{Y_{I,\alpha}\}$. 
The width of the fluctuation along each of $9N^2$ directions is of order $N^0$. Therefore, the location of the center of the wave packet can be distinguished from the origin if $\sqrt{\sum_I{\rm Tr}Y_I^2}$ is sufficiently larger than $1$.
If there are two wave packets around $\{Y_{I,\alpha}\}$ and $\{Y'_{I,\alpha}\}$, they can be distinguished if $\sqrt{\sum_I{\rm Tr}(Y_I-Y'_I)^2}$ is sufficiently larger than $1$.


The problem with the past treatment~\cite{Susskind:1998vk,Polchinski:1999br} was that they took a typical configuration in the path integral, or a typical result in the measurement of $\hat{X}_{I,ij}$, and diagonalized it without separating the slow modes, that can actually be diagonalized, and the fast modes, that cannot really be diagonalized.\footnote{
To their credit, they clearly pointed out the necessity of the separation of slow and fast modes, but did not identify a concrete procedure for the separation.}
A better procedure is to diagonalize the center of the wave packet, or equivalently, 
to diagonalize the expectation value of $\hat{X}$. 
This procedure has a \textit{well-defined} meaning at the level of the quantum states in the Hilbert space.

Another way to phrase it is that the past treatment was the gauge fixing of $|X\rangle$ rather than that of $|{\rm ground\ state}\rangle$ or $|{\rm wave\ packet\ at\ } \{Y_{I,\alpha}\}\rangle$. The position-eigenstate $|X\rangle$ is not the low-energy state relevant for physics under consideration; the uncertainty principle forces us to consider a wave packet.

Let us see a few special cases whose meanings are obvious.
\begin{itemize}
\item
Let us separate one of the D-branes from others sitting at the origin, without exciting any open string. 
Specifically, we can construct a wave packet centered around a point $\vec{Y}_{ij}=\delta_{iN}\delta_{jN}\vec{y}\in\mathbb{R}^{9N^2}$, as
\begin{align}
e^{-i\vec{y}\cdot\hat{\vec{P}}_{NN}}|{\rm ground\ state}\rangle. 
\label{eq:N-th-D-brane-separated}
\end{align} 
As long as $|\vec{y}|\gtrsim 1$, the position of the probe is a legitimate notion.  

\item
By using the U(1)-part we can easily make a U($N$)-invariant state describing $N$-coincident D0-branes at point $\vec{y}$, as
\begin{align} 
e^{-i\vec{y}\cdot(\sum_{k=1}^N\hat{\vec{P}}_{kk})}|{\rm ground\ state}\rangle. 
\label{N-coincident-D-brane}
\end{align}
Note that this full U($N$)-invariance is exactly what we expect when $N$ D-branes are sitting on top of each other~\cite{Witten:1995im}.

\item
We can construct a state describing `diagonal matrices' $\vec{Y}_{ij}=\vec{y}_i\delta_{ij}$ as 
\begin{align} 
e^{-i\sum_{k=1}^N\vec{y}_k\cdot\hat{\vec{P}}_{kk}}|{\rm ground\ state}\rangle. 
\end{align}
If some $\vec{y}_i$'s take the same value, say $N_1$ of them are $\vec{x}$, $N_2$ of them are $\vec{x}'$ and so on, then such a state is invariant under U($N_1$)$\times$U($N_2$)$\times\cdots$. 
This symmetry enhancement is consistent with the interpretation that $\vec{y}_i$ is the location of $i$-th D-brane~\cite{Witten:1995im}. 

\end{itemize}

We can add further justification for the interpretation that the center of the wave packet should be identified with the `location of D-branes', and more generally, `matrices'.\footnote{
The author would like to thank Alexey Milekhin for useful discussion regarding this point.
}  
The Hamiltonian $\hat{H}$ is a polynomial of $\hat{P}_I$ and $\hat{X}_I$, so let us write it as $\hat{H}=H(\hat{P},\hat{X})$. 
Then we can show that 
\begin{align}
e^{i\sum_I{\rm Tr}(Y_I\hat{P}_I)}
H(\hat{P},\hat{X})
e^{-i\sum_I{\rm Tr}(Y_I\hat{P}_I)}
=
H(\hat{P},\hat{X}+Y). 
\label{eq:background-field}
\end{align}
Therefore, instead of acting $H(\hat{P},\hat{X})$ on $|{\rm wave\ packet\ at\ } \{Y_{I,\alpha}\}\rangle$, 
we could act $H(\hat{P},\hat{X}+Y)$ on $|{\rm ground\ state}\rangle$, if we like.  
In the latter treatment, when the coupling constant $g^2$ is nonzero (which will be studied in Sec.~\ref{sec:MM-finite-coupling}), if we take $Y$ to be diagonal, 
the mass terms for the off-diagonal elements are generated from the commutator-squared term in the action.
They are identified with the mass terms for open strings~\cite{Witten:1995im}.

We emphasize that the coherent state discussed here is just \textit{one of many possible realizations} of the wave packets. 
When the interaction is introduced, 
it may or may not be a stable wave packet. 
If we consider strongly-coupled theories with gravity duals, D-brane probes in the gravity side may not be described by the coherent state precisely, and a large modification may be needed. 
We will discuss this point further in later sections.

\begin{figure}[htbp]
  \begin{center}
   \includegraphics[width=100mm]{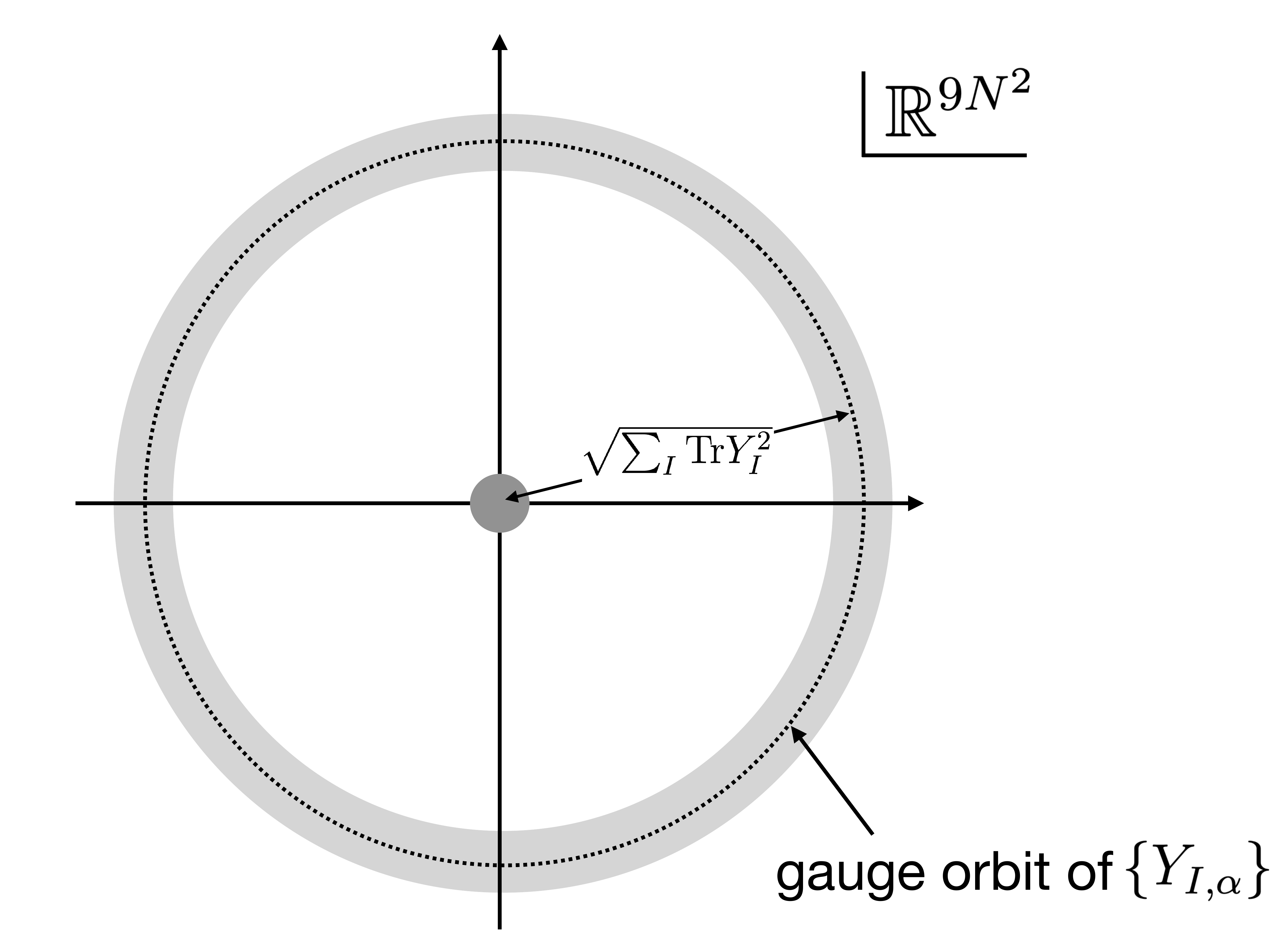}
  \end{center}
  \caption{
  Gauge-symmetrized version of the wave packet shown in Fig.~\ref{fig:wave-packet} (gray ring) . 
  The wave function is localized near the gauge orbit of $\{Y_{I,\alpha}\}$ (dotted circle) . 
  The ground state (gray disk) is localized around the origin.  
  Note that the shape and volume of the gauge orbit depends on $\{Y_{I,\alpha}\}$.
     }\label{fig:wave-packet-gauge-inv}
\end{figure}

Note also that, generically, these states are not U($N$)-invariant, that is, they belong to ${\cal H}_{\rm ext}$ but not to ${\cal H}_{\rm inv}$. If we want to discuss everything in terms of ${\cal H}_{\rm inv}$, 
we have to project them to the singlet sector. 
Equivalently, we can take a superposition of all wave packets along the gauge orbit;\footnote{
In \eqref{eq:wave-packet-generic-with-momentum},
we should choose $Q_I$ appropriately such that 
the configuration does not move along the gauge orbit. }$^,$\footnote{
An implicit but important assumption here is that $|\Phi\rangle$ and its U($N$)-symmetrized version have similar properties, 
in that the expectation values of gauge-invariant operators are identical up to small corrections.}
see Fig.~\ref{fig:wave-packet-gauge-inv}.
On the other hand, the ground state is automatically gauge-invariant without performing such a projection. 
In this sense, the ground state is `genuinely' U($N$)-invariant~\cite{Hanada:2020uvt}. The $N$-coincident-D-brane state \eqref{N-coincident-D-brane} is also genuinely U($N$)-invariant. In Appendix~\ref{sec:BEC}, we explain how such `genuine' gauge singlets can be distinguished from the other kind of singlets.

The same situation appears in a system of $N$ identical bosons, which can be regarded as a gauged quantum mechanics of $N$-component vectors~\cite{Hanada:2020uvt}. That the bosons are `identical' means the physical states have to be invariant under the S$_N$ permutation, hence this system is a gauge theory with S$_N$ gauge group. This system can be analyzed by using the extended Hilbert space, and Bose-Einstein condensation~\cite{einstein1924quantentheorie} is characterized by the same `genuine' gauge invariance~\cite{feynman_superfluidity1,feynman_superfluidity2,feynman_superfluidity3,PhysRev.104.576,RevModPhys.34.694}.

\subsubsection{Generic excited states}
\hspace{0.51cm}
Let us take a generic wave packet, by taking $Y_I$ and $Q_I$ in \eqref{eq:wave-packet-generic-with-momentum} to be generic matrices whose eigenvalues are of order $\sqrt{N}$. 
(More generally, we can take a superposition of such states.)
Generically, such a state is not invariant under any U($N$) transformation \eqref{gauge-transf-Fock-state}, except that any state is trivially invariant under the adjoint action of the U($1$) part. Therefore we can choose a gauge such that the diagonal elements are more highly excited than the off-diagonal elements. In this case the size of the bound state is actually of order $\sqrt{N}$. 

The same holds for other kinds of excited states such as the Fock state $|\{n_{I,\alpha}\}\rangle\in{\cal H}_{\rm ext}$ at sufficiently high energy.
\subsubsection{Partially-deconfined states}\label{sec:partial-deconfinement}
\hspace{0.51cm}
More interesting physics can be observed at the intermediate energy scale. 
As shown in Ref.~\cite{Hanada:2019czd}, there are two phase transitions at\footnote{
We subtracted the zero-point energy $\frac{9N^2}{2}$.
} $\frac{E}{N^2}=0$ (Hagedorn transition) and $\frac{E}{N^2}=\frac{1}{4}$ (Gross-Witten-Wadia transition).
In between these two phase transitions, at $E=\frac{M^2}{4}$, the U($M$) subgroup of U($N$) is deconfined. 
This is a particular example of partial deconfinement~\cite{Hanada:2016pwv,Berenstein:2018lrm,Hanada:2018zxn,Hanada:2019czd,Watanabe:2020ufk}
that is conjectured to be a generic feature among various large-$N$ gauge theories.
We can fix a gauge such that deconfinement is taking place in the $M\times M$ upper-left block.
Equivalently, we restrict $Y_I$ and $Q_I$ to be $M\times M$. 
This fixes U($N$) down to U($M$)$\times$U($N-M$).
We can further fix U($M$) such that the diagonal entries of the deconfined block becomes as large as $O(\sqrt{M})$.
The `genuine' symmetry U($N-M$) is left unfixed. 
Hence we obtain a bound state whose radius is $\sim\sqrt{M}$.
This bound state is conjectured to be the gauge-theory realization of the small black hole~\cite{Hanada:2016pwv}. (See also Ref.~\cite{Krishnan:2020oun} for a recent application of this idea to the black hole evaporation.)

As a probe, we can excite the $(N,N)$ component. 
The notion of a `location' can make sense if the distance from the origin is sufficiently larger than the `BH radius' $\sim\sqrt{M}$.

\subsection{Finite coupling}\label{sec:MM-finite-coupling}
\hspace{0.51cm}
Even at finite coupling ($g^2>0$), we can expect that the confining vacuum is `genuinely' gauge-invariant, even in the extended Hilbert space containing the non-singlet modes.
While this is a natural assumption, some of the audience would request the evidence. 
For small system size, we can check it numerically. In the large-$N$ limit, the distribution of the phases of the Polyakov loop can be used to see if a given state in ${\cal H}_{\rm inv}$ is `genuinely' gauge-invariant~\cite{Hanada:2020uvt}. 
As a starting point, let us write the canonical partition function at temperature $T$ as
\begin{align}
Z(T)=\frac{1}{{\rm volU}(N)}\int dU{\rm Tr}_{{\cal H}_{\rm ext}}(\hat{U}e^{-\hat{H}/T}), 
\end{align}
where ${\rm Tr}_{{\cal H}_{\rm exit}}$ is the trace in the extended Hilbert space. 
Here $U$ is an element of U($N$), and the integral is taken by using the Haar measure. 
The operator $\hat{U}$ enforces the gauge transformation. 
This $U$ corresponds to the Polyakov line in the path integral formulation. 
The contribution of the ground state is 
\begin{align}
\frac{1}{{\rm volU}(N)}
\int dU e^{-E_0/T}\langle {\rm ground\ state}|\hat{U}|{\rm ground\ state}\rangle, 
\end{align} 
where $E_0$ is the energy of the ground state. 
If the ground state is not genuinely U($N$)-invariant, there are degenerate vacua in ${\cal H}_{\rm ext}$ related by gauge transformation, and we need to sum them up. Either way, only such $U\in{\rm U}(N)$ that leaves $|{\rm ground\ state}\rangle$ invariant can contribute to the partition function.
This $U$ is the Polyakov loop. That the phases of the Polyakov loop is uniform at zero temperature is consistent with the genuine U($N$)-invariance of the ground state, i.e., it is invariant under any U($N$) transformation. For details, see Ref.~\cite{Hanada:2020uvt}. 
Note that this argument is essentially identical to the characterization of Bose-Einstein condensation of $N$ indistinguishable bosons via the S$_N$-invariance~\cite{feynman_superfluidity1,feynman_superfluidity2,feynman_superfluidity3}.
 
In the 't Hooft large-$N$ limit ($\lambda=g^2N\sim N^0$, $T\sim N^0$, $E\sim N^2$), 
the expectation value $\langle{\rm Tr}X_I^2\rangle$ is proportional to $N^2$. 
At low temperature $T\ll\lambda^{1/3}$ and strong coupling $\lambda\gg 1$, it scales as $\langle{\rm Tr}X_I^2\rangle\sim \lambda^{-1/3}N^2$.   
From this, in the past it has been interpreted that the size of the ground state wave function is $\lambda^{-1/6}N^{1/2}$. 
However, with a natural assumption that the ground-state wave function is gauge-invariant, 
this scaling simply means that the width of the ground-state wave function is proportional to $\lambda^{-1/6}$ with respect to each direction of $\mathbb{R}^{9N^2}$. 
Just as in the free theory, we can introduce a probe by exciting the $(N,N)$ component by using \eqref{eq:N-th-D-brane-separated}, 
by taking $|{\rm ground\ state}\rangle$ to be the vacuum of the interacting theory. 
Such probe is well outside the bound state of $N-1$ D-branes if the distance from the origin is sufficiently larger than $\lambda^{-1/6}$.  
We can use \eqref{eq:wave-packet-generic} to construct various other wave packets.


\subsubsection*{Correction to the coherent state}
\hspace{0.51cm}
As explained in the paragraph that contains \eqref{eq:background-field},
the open-string mass term is naturally induced by considering a wave packet \eqref{eq:wave-packet-generic}.
For example, if we put the $(N,N)$ component at $\vec{y}=(L,0,0,\cdots,0)$, 
then the induced mass term is 
$g^2L^2\sum_{I=2}^9\sum_{i=1}^{N-1}|\hat{X}_{I,iN}|^2$. 
%
A caveat here is that we did not touch the off-diagonal elements. 
If the off-diagonal elements acquire mass due to the Higgsing, 
the energy of the wave packet becomes large unless 
the width of the wave packet along these directions in $\mathbb{R}^{9N^2}$ 
(in the example above, $\hat{X}_{I,iN}$, $I=2,\cdots,8$, $i=1,2,\cdots,N-1$) shrink.
We did not take into account such effects.
In order to obtain a stable, low-energy wave packet, probably we should fix the location of the center of the wave packet and then minimize the energy:\footnote{
It would be better to use the U($N$)-symmetrized version of $|\Phi\rangle$ (Fig.~\ref{fig:wave-packet-gauge-inv}) to evaluate the energy. 
}
\begin{itembox}[c]{A natural construction of wave packet $|\Phi\rangle$ at finite coupling}
Minimize $\langle\Phi|\hat{H}|\Phi\rangle$ satisfying the constraints $\langle\Phi|\hat{X}_{I}|\Phi\rangle=Y_I$ and $\langle\Phi|\hat{P}_{I}|\Phi\rangle=Q_I$.
\end{itembox}
The symmetry enhancement we observed for the coherent states in Sec.~\ref{sec:coherent-state} persists here:
if $(Y_I,Q_I)$ is invariant under the action of a subgroup of U($N$), 
the corresponding quantum state $|\Phi\rangle$ is also invariant. 

\subsubsection*{Partially-deconfined states}
\hspace{0.51cm}
The strong coupling limit ($\lambda\to\infty$) has been studied numerically via lattice Monte Carlo simulation and partial deconfinement has been demonstrated~\cite{Watanabe:2020ufk,Bergner:2019rca}. 
Therefore, the argument provided in Sec.~\ref{sec:partial-deconfinement} can be repeated. 
The deconfined sector in the partially-deconfined state is interpreted as a thermally-excited bound state that is analogous to the small black hole in string theory. 

\subsection{D0-brane matrix model}\label{sec:D0-MM}
\hspace{0.51cm}
The argument given above applies to the D0-brane matrix model~\eqref{action:BFSS} as well.\footnote{
We may have to remove the flat direction in order to pick up the gauge-invariant vacuum. It can be achieved e.g., by adding a small mass to scalars, take the large-$N$ limit and then removing the mass. 
} 
Modulo a natural assumption that the ground state is genuinely gauge-invariant, 
the scaling $\langle{\rm Tr}X_I^2\rangle\sim \lambda^{-1/3}N^2$ at low temperature simply means that the width of the ground-state wave function with respect to each direction of $\mathbb{R}^{9N^2}$ is $\lambda^{-1/6}$.

We repeat that the coherent state \eqref{eq:wave-packet-generic}, and more generally \eqref{eq:wave-packet-generic-with-momentum}, is merely \textit{one of many possible realizations} of the wave packet.
An apparent issue when we try to relate the coherent state to the probe D-brane in gravity side is supersymmetry: the wave packet ought to be supersymmetric when $Y_I$'s are simultaneously diagonalizable and $Q_I$'s vanish. 
Probably the most natural wave packet $|\Phi\rangle$ is obtained by minimizing $\langle\Phi|\hat{H}|\Phi\rangle$ with the constraints $\langle\Phi|\hat{X}_{I}|\Phi\rangle=Y_I$ and $\langle\Phi|\hat{P}_{I}|\Phi\rangle=Q_I$; see Sec.~\ref{sec:MM-finite-coupling}.
After the gauge transformation, the wave packet is localized about $Y_I^{(U)}$ and $Q_I^{(U)}$.

We expect that the D0-brane matrix model has significantly richer dynamics than the bosonic theories. 
Dual gravity analysis of this system~\cite{Itzhaki:1998dd}, combined with the analogy to the partial-deconfinement proposal for 4d SYM~\cite{Hanada:2016pwv}, leads to the following \textit{speculations} regarding finite-temperature physics:
\begin{itemize}
\item
At $\lambda^{1/3}N^{-5/9}\lesssim T\ll\lambda^{1/3}$, the system is dual to black zero-brane in type IIA superstring theory~\cite{Itzhaki:1998dd}.   
The black zero-brane is analogous to the large black hole in AdS which has positive specific heat. 
According to the proposal in Ref.~\cite{Hanada:2016pwv,Hanada:2018zxn}, such states should be completely-deconfined.
(Still, at very low energy, the off-diagonal elements should be highly suppressed; otherwise the energy cannot be parametrically small. In this sense, this state may be almost block-diagonal, and the size of the block shrinks at low temperature. See Appendix~\ref{sec:gauged-vs-ungauged} for a related material.)
If we simply identify the size of the bound state $R$ and the radius of black zero-brane, 
we obtain $R\sim(\lambda T^2)^{1/5}N^{1/2}$. 

\item
As the energy goes down, the finite extent of the M-theory circle becomes non-negligible. 
Around $T\sim \lambda^{1/3}N^{-5/9}$, the transition to eleven-dimensional Schwarzschild black hole takes place~\cite{Itzhaki:1998dd}. 
Below this energy scale, the specific heat is negative, i.e., temperature goes up as the energy goes down and black hole shrinks.
Such phase is naturally described by partially-deconfined states~\cite{Hanada:2016pwv,Hanada:2018zxn}.\footnote{
There is a subtle difference from the original proposal~\cite{Hanada:2016pwv}: 
in the original proposal $N-M$ D-branes not contributing to black hole were supposed to be hovering somewhere outside black hole, but in the current proposal they are sitting at the center of the bulk. 
The same holds for a proposal on the small black hole in AdS$_5\times$S$^5$ discussed in Sec.~\ref{sec:4dSYM}.
}$^,$\footnote{
The idea that a nontrivial $M$-dependence may explain the negative specific heat of the eleven-dimensional Schwarzschild black hole was proposed in Ref.~\cite{Berkowitz:2016znt}, though that reference contains a few apparent mistakes and confusions. 
}
It would be natural to identify the size of the $M\times M$ deconfined block with the radius of the Schwarzschild black hole. 

\end{itemize}

In principle, these speculations can be tested via lattice Monte Carlo simulation,\footnote{See Ref.~\cite{Watanabe:2020ufk} for the analysis of the partially-deconfined phase in the bosonic matrix model.} or perhaps also via the machine-learning approach along the line of Ref.~\cite{Han:2019wue}. 

By generalizing the probe picture, and by following the philosophy of the Matrix Theory conjecture~\cite{Banks:1996vh}, it would be natural to interpret the small bound states as physical objects such as a graviton or tiny black hole. For example, a simple operator 
\begin{align}
{\rm Tr}(\hat{X}_I\hat{X}_J\hat{X}_K)=\sum_{p,q,r=1}^N\hat{X}_I^{pq}\hat{X}_J^{qr}\hat{X}_K^{rp}
\end{align}
is the U($N$)-symmetrized version of 
\begin{align}
\hat{X}_I^{N,N-1}\hat{X}_J^{N-1,N-2}\hat{X}_K^{N-2,N}, 
\end{align}
and hence it can be regarded as a small bound state occupying a $3\times 3$ block. Machine learning~\cite{Han:2019wue} and quantum simulation~\cite{Gharibyan:2020bab} can be be practically useful approach to study such small bound state.

\section{$(3+1)$-d Yang-Mills}\label{sec:4dSYM}
\hspace{0.51cm}
The same puzzle regarding the size of the bound state existed for quantum field theories including $(3+1)$-d maximal SYM compactified on S$^3$ (see e.g., Ref.~\cite{Heemskerk:2012mn}). The resolution provided for the matrix models can work for quantum field theories as well, because the key ingredient --- `genuine' gauge-invariance of the ground state --- is not specific to the matrix models. The only difference is that D3-brane can take a nontrivial shape, namely $X_{I,ij}$ can be a nontrivial function on S$^3$.  

The weak-coupling limit of $(3+1)$-d maximal Yang-Mills on S$^3$ can be studied analytically via technologies introduced in Refs.~\cite{Aharony:2003sx,Sundborg:1999ue}, regardless of the details of the theory such as supersymmetry or matter content.
We can explicitly confirm the `genuine' gauge-invariance of the ground state~\cite{Hanada:2020uvt} and partial deconfinement in the intermediate-energy regime~\cite{Hanada:2019czd} with the size of the U($M$)-deconfined states scaling as $\sqrt{M}$.

Strong coupling region is challenging, nonetheless let us make a crude, heuristic estimate.
(The following is essentially the argument in Ref.~\cite{Hanada:2016pwv}, with slight improvement.) 
For concreteness, we take the radius of $S^3$ to be $R_{{\rm S}^3}=1$. 
We use the normalization ${\cal L}=\frac{1}{4g^2}{\rm Tr}\left(F_{\mu\nu}^2+\cdots\right)$, in which the 't Hooft counting is straightforward. 

Our hypothesis is that the thermal bound state (deconfined block) is a black hole, and we identify the radius of the thermal bound state with the radius of black hole up to a multiplicative factor.
Hence let us first estimate the radius of the thermal bound state.
We focus on the U($M$)-partially-deconfined state, and assume\footnote{
This is a highly nontrivial assumption, given that we are studying the strongly-coupled region. 
} that the radius and the energy of the thermal bound state can roughly be estimated by truncating $N\times N$ matrices to $M\times M$, 
with the effective 't Hooft coupling $\lambda_M\equiv g^2M$.   
This truncated system is strongly coupled when $\lambda_M\gg 1$, and there the interaction term $\frac{1}{g^2}{\rm Tr}[X_I,X_J]^2=\frac{M}{\lambda_M}{\rm Tr}[X_I,X_J]^2$ dominates the dynamics. By noticing that the dependence on $\lambda_M$ disappears when $\tilde{X}_I\equiv \lambda_M^{-1/4}X_I$ is used, we can see that the eigenvalues of $\tilde{X}_I$ are of order $M^0$, and those of $X_I$ are proportional to $\lambda_M^{1/4}$.\footnote{
Here by the `eigenvalues' we mean the slow-mode contribution. 
}
Hence we estimate that the radius of black hole $R_{\rm BH}$ is proportional to $M^{1/4}$.  
In our setup $R_{\rm BH}$ is of order 1 when the transition between large and small black holes takes place, and this transition should be at  $M\sim N$. 
Therefore,  $R_{\rm BH}\sim\left(\frac{M}{N}\right)^{1/4}$, and $T_{\rm BH}\sim\left(\frac{M}{N}\right)^{-1/4}$. 
  
The next step is to estimate the entropy $S_{\rm BH}$. 
From the 't Hooft counting, the entropy $S_{\rm BH}$ should be written as $S_{\rm BH}\sim f(\lambda_M)\cdot M^2$, with some function $f$. 
To determine $f$, we again look at the transition between large and small black holes takes place. There the entropy is simply proportional to $N^2$ as long as $\lambda=g^2N$ is large, and hence, we conclude $f(\lambda_M)$ is just constant at $\lambda_M\gg 1$, and the entropy is $S_{\rm BH}\sim M^2$.  

By combining
$R_{\rm BH}\sim\left(\frac{M}{N}\right)^{1/4}$,  
$T_{\rm BH}\sim\left(\frac{M}{N}\right)^{-1/4}$
and 
$S_{\rm BH}\sim M^2$, 
we obtain 
\begin{align}
S_{\rm BH}\sim N^2R_{\rm BH}^{8}\sim\frac{R_{\rm BH}^8}{G_{\rm N}}\sim\frac{1}{G_{\rm N}T_{\rm BH}^8},
\end{align} 
where $G_{\rm N}$ is the ten-dimensional Newton constant. 
This is the expected behavior of the small black hole.  

If the effective coupling describing the thermal bound state is small ($g^2M\lesssim 1$), it can be described in terms of long free strings. 
Strong-coupling description (small black hole) and weak-coupling description (free string) should be switched at $g^2M\sim 1$, 
that translates to $S_{\rm BH}\sim g^{-4}$. This is the same as the expectation from the dual gravity analysis~\cite{Aharony:1999ti,Aharony:2003sx}. 

This argument is  based on many nontrivial assumptions (including that partial deconfinement takes place at the strongly-coupled region of 4d SYM), and hence we do not claim it is a `derivation'. 
Our purpose here was to show how the bulk geometry, including black hole, might be described by color degrees of freedom. 
A better test might be doable by using the index~\cite{Kinney:2005ej} with complex chemical potential~\cite{Choi:2018vbz}.

\section{Future Directions}\label{sec:future-directions}
\hspace{0.51cm}

In this paper we suggested that a classic way of seeing the emergent bulk geometry, analogous to the Matrix Theory proposal by Banks, Fischler, Shenker and Susskind~\cite{Banks:1996vh} --- roughly speaking, `eigenvalues are coordinates' --- can make sense in the Maldacena-type gauge/gravity duality~\cite{Maldacena:1997re,Itzhaki:1998dd}. 
The key was to understand the meaning of `matrices' and `eigenvalues' precisely.
Because we are interested in low-energy states, we need to consider a wave packet whose center is identified with `matrices'. 
The genuine gauge invariance of the ground state~\cite{Hanada:2020uvt} played the important role for the determination of the size of the ground-state wave function.

A natural expectation would be that probe D-branes, whose locations are identified with the diagonal elements of the `matrices', are described by the Dirac-Born-Infeld action in the black-brane spacetime as in Maldacena's original proposal~\cite{Maldacena:1997re}. 
(Note however that the determination of the appropriate wave packet is a nontrivial problem that requires further study, as we emphasized a few times in this paper.)
An ideal way to test it is to realize supersymmetric gauge theories on a quantum computer~\cite{Gharibyan:2020bab} and then perform the D-brane-scattering experiments. See Refs.~\cite{Danielsson:1996uw,Kabat:1996cu,Douglas:1996yp,Becker:1997xw,Okawa:1998pz} for analytic calculations related to such scattering processes. 
Another interesting approach is the machine-learning method to obtain the wave function~\cite{Han:2019wue}, which might be useful for determining the potential energy as a function of the location of the probe D-brane.
Such an approach is analogous to the analysis based on the probe effective action via path integral~\cite{Jevicki:1998ub}.
Refs.~\cite{Ferrari:2012nw,Ferrari:2013aba} propose a way of detecting the emergent space by using the probe action, that may be connected our proposal. 
Monte Carlo simulation based on Euclidean path integral can also be a powerful tool. 
In the past, similar but slightly different setups were studied.  
In Ref.~\cite{Rinaldi:2017mjl} the $(N,N)$ component was Higgsed by adding an extra term to the potential, and the interaction between the probe and thermal bound state was studied. The parameter region studied in that paper was $T\gtrsim\lambda^{1/3}$, where the subtlety associated with `delocalization' in the path-integral picture was not the important issue. 
Ref.~\cite{Filev:2015cmz} used the D0/D4-system described by the Berkooz-Douglas matrix model~\cite{Berkooz:1996is}.  
A theoretically cleaner setup would be to use
\begin{align}
Z(T;Y)
&\equiv
\sum_{|E\rangle\in{\cal H}_{\rm inv}}
\langle E|
e^{-H(\hat{P},\hat{X}+Y)/T}
|E\rangle 
\end{align}
to estimate the interaction between the probe and black hole, 
by using a coherent state as a probe.
Although the coherent state may not be an ideal probe, 
there may be a qualitative change when it goes into the thermal bound state.

Another interesting direction is to understand the relationship to other approaches to the emergent space. 
This is very important toward the understanding of the interior of the black hole, where a simple geometric picture discussed in this paper may not be applicable.
Recently there are several attempts to use the entanglement between color degrees of freedom for this purpose~\cite{Mazenc:2019ety,Das:2020xoa,Das:2020jhy,Hampapura:2020hfg,Anous:2019rqb,Alet:2020ehp,Mollabashi:2014qfa}. 
It would be useful to study the meanings of these proposals, or to make a better proposal, based on the geometric picture discussed in this paper. 
For example, for the D0-brane quantum mechanics, we can consider a wave packet localized about $\vec{Y}_{ij}=\vec{y}_{i}\delta_{ij}$, where $\vec{y}_1,\cdots,\vec{y}_M\in A\subset\mathbb{R}^9$ and 
$\vec{y}_{M+1},\cdots,\vec{y}_N\in\bar{A}\subset\mathbb{R}^9$. 
Then we can integrate out the upper-left $M\times M$ block to define the entanglement entropy between the probes in a region $A$ and those in a region $\bar{A}$. 

How can we see the `shape' of a bound state? One natural approach is to make it `maximally diagonal', for example by fixing $U\in{\rm U}(N)$ such that 
$\sum_{I=1}^9\sum_{i=1}^N|(UX_IU^{-1})_{ii}|^2$ is maximized~\cite{Azeyanagi:2009zf,Hotta:1998en}.
In the past this procedure was applied by using typical configurations in the path integral as `matrices'.  
Obviously, we should apply this procedure to the slow modes. 

The IKKT matrix model~\cite{Ishibashi:1996xs} is another interesting model that may exhibit the emergence of spacetime. 
It is more ambitious than the class of theories discussed in this paper, in that even time direction should emerge from color degrees of freedom. 
The argument in this paper does not apply to the IKKT matrix model because we assumed the existence of time when we defined the Hamiltonian.  
It would be interesting to think about a proper definition of `diagonalization' and `eigenvalue distribution' in this model. 

\begin{center}
\textbf{Acknowledgement}\\
\end{center}
\hspace{0.51cm}
The author would like to thank 
D.~Anninos, T.~Anous, S.~Das, S.~Funai, X.~Han, G.~Ishiki, J.~Maldacena, G.~Mandal, S.~Matsuura, A.~Milekhin, E.~Rinaldi, H.~Shimada, B.~Swingle, S.~Trivedi, T.~Wiseman and T.~Yoneya for discussions and comments. 
He thanks the International Centre for Theoretical Sciences (ICTS) for hosting the online program ``Nonperturbative and Numerical Approaches to Quantum Gravity, String Theory and Holography" (code: ICTS/numstrings2021/1), which gave him a valuable opportunity of discussing the materials presented in this paper with several participants. 
He was supported by the STFC Ernest Rutherford Grant ST/R003599/1.

\appendix
\section{Relation between Bose-Einstein Condensation (BEC) and color confinement, and `genuine' gauge invariance}\label{sec:BEC}
\hspace{0.51cm}

\subsection{${\cal H}_{\rm ext}$ and ${\cal H}_{\rm inv}$}\label{app:Hinv-vs-Hext}
\hspace{0.51cm}
Let us consider generic gauge group $G$. 
The canonical partition function of gauge theory is defined as 
\begin{align}
Z(T)={\rm Tr}_{{\cal H}_{\rm inv}}(e^{-\hat{H}/T}). 
\label{partition-function-Hinv}
\end{align}
Let us show that this can also be written as 
\begin{align}
Z(T)=\frac{1}{{\rm vol}(G)}\int_G dg{\rm Tr}_{{\cal H}_{\rm ext}}(\hat{g}e^{-\hat{H}/T}), 
\label{partition-function-Hext}
\end{align}
where ${\rm vol}(G)$ is the volume of $G$.

Let $|\Phi\rangle\in{\cal H}_{\rm ext}$ be an energy eigenstate in certain gauge. 
It can be projected to a singlet state $|\Phi\rangle_{\rm inv}\in{\cal H}_{\rm inv}$ as
\begin{align}
|\Phi\rangle_{\rm inv}
=
\frac{1}{\sqrt{C_\Phi}}\int_G dg \left(\hat{g}|\Phi\rangle\right), 
\end{align}
where the integral is taken over the gauge group $G$ by using the Haar measure, and $\hat{g}$ generates the gauge transformation associated with the group element. 
The normalization factor $C_\Phi$ is 
\begin{align}
C_\Phi
=
\int_G dg\int_G dg' \left(\langle\Phi|\hat{g}^{-1}\right)\left(\hat{g}'|\Phi\rangle\right)
=
{\rm vol}(G)\cdot
\int_G dg\langle\Phi|\hat{g}|\Phi\rangle
=
{\rm vol}(G)\cdot{\rm vol}(G_\Phi), 
\end{align}
where $G_\Phi$ is a subgroup of $G$ that leaves $|\Phi\rangle$ invariant. 

When the trace is taken over the extended Hilbert space, the over-counting factor associated with an energy eigenstate $|\Phi\rangle$ is $\frac{{\rm vol}(G)}{{\rm vol}(G_\Phi)}$.
Therefore, 
\begin{align}
{\rm Tr}_{{\cal H}_{\rm inv}}(e^{-\hat{H}/T})
=
\sum_{\Phi}\frac{{\rm vol}(G_\Phi)\cdot e^{-E_\Phi/T}}{{\rm vol}(G)}, 
\end{align}
where the sum with respect to energy eigenstates $|\Phi\rangle$ is taken over ${\cal H}_{\rm ext}$. 
We can also show that
\begin{align} 
 \int_G dg{\rm Tr}_{{\cal H}_{\rm ext}}(\hat{g}e^{-\hat{H}/T})
 =
  \int_G dg
 \sum_{\Phi}e^{-E_\Phi/T}
 \langle\Phi|\hat{g}|\Phi\rangle
 =
 \sum_{\Phi}{\rm vol}(G_\Phi)\cdot e^{-E_\Phi/T}. 
 \label{Z-generic-G}
\end{align} 
Therefore, 
\eqref{partition-function-Hinv} and \eqref{partition-function-Hext} are equivalent.

\subsection{BEC, confinement and `genuine' gauge invariance}
\hspace{0.51cm}
We emphasized the importance of the `genuine' gauge invariance throughout this paper. 
A crisp characterization of this notion can be illuminated via the close connection between Bose-Einstein condensation and color confinement at large $N$~\cite{Hanada:2020uvt}.

Let us consider a system of $N$ free bosons in the harmonic trap. The Hamiltonian is 
\begin{align}
\hat{H}
=
\sum_{i=1}^N
\left(
\frac{\hat{\vec{p}}_i^2}{2m}
+
\frac{m\omega^2}{2}\hat{\vec{x}}_i^2
\right), 
\end{align}
where $\hat{\vec{x}}_i=(\hat{x}_i,\hat{y}_i,\hat{z}_i)$ and $\hat{\vec{p}}_i=(\hat{p}_{x,i},\hat{p}_{y,i},\hat{p}_{z,i})$ are the coordinate and momentum of $i$-th particle. 

Because $N$ bosons are indistinguishable, this is a gauged quantum mechanics with the gauge group S$_N$.
As the basis of the extended Hilbert space ${\cal H}_{\rm ext}$, 
we can use the Fock states $|\vec{n}_1,\cdots,\vec{n}_N\rangle$, which are energy eigenstates with the energy 
$E=\sum_{i=1}^NE_{\vec{n}_i}=\sum_{i=1}^N\left((n_{x,i}+n_{y,i}+n_{z,i})\omega+\frac{3}{2}\right)$.  
The partition function is given by
\begin{align}
Z(T)
&=
\frac{1}{N!}
\sum_{\sigma\in{\rm S}_N}
\sum_{\vec{n}_1,\cdots,\vec{n}_N}
\langle\vec{n}_1,\cdots,\vec{n}_N|
\hat{\sigma}e^{-\hat{H}/T}
|\vec{n}_1,\cdots,\vec{n}_N\rangle
\nonumber\\
&=
\frac{1}{N!}
\sum_{\vec{n}_1,\cdots,\vec{n}_N}
e^{-(E_{\vec{n}_1}+\cdots+E_{\vec{n}_N})/T}
\left(
\sum_{\sigma\in{\rm S}_N}
\langle\vec{n}_1,\cdots,\vec{n}_N
|\vec{n}_{\sigma(1)},\cdots,\vec{n}_{\sigma(N)}\rangle
\right). 
\label{BEC-partition-function}
\end{align}
The factor 
$\sum_{\sigma\in{\rm S}_N}
\langle\vec{n}_1,\cdots,\vec{n}_N
|\vec{n}_{\sigma(1)},\cdots,\vec{n}_{\sigma(N)}\rangle$
counts the number of $\sigma\in{\rm S}_N$ that leaves $|\vec{n}_1,\cdots,\vec{n}_N\rangle$ invariant. 
(This is corresponds to ${\rm vol}(G_\Phi)$ in \eqref{Z-generic-G}.)
If all $\vec{n}_i$'s are the same (e.g., the ground state, $\vec{n}_1=\cdots\vec{n}_N=\vec{0}$), then a large enhancement factor $N!$ appears. Let us call such states `genuinely S$_N$-invariant states'. 
For generic states most permutations $\sigma\in{\rm S}_N$ change the state and hence such enhancement factor does not appear. 

Bose-Einstein condensation~\cite{einstein1924quantentheorie} is a phenomenon that many particles fall into the ground state. 
It is triggered by the enhancement mechanism mentioned above: if $M$ particles are excited while $N-M$ particles are in the ground state, then the enhancement factor $(N-M)!$ appears from the latter.  
The same mechanism triggers color confinement: ${\rm vol}(G_\Phi)$ in \eqref{Z-generic-G} serves as the enhancement factor, 
and genuinely gauge-invariant state that satisfies $G=G_\Phi$ becomes dominant at low temperature. 
Partially-BEC phase corresponds to the partially-confined phase ($=$ partially-deconfined phase). 

As we mentioned in Sec.~\ref{sec:MM-finite-coupling}, that the distribution of the Polyakov loop in U($N$) gauge theory becomes uniform at low temperature is the consequence of the genuine gauge invariance of the ground state. 
Exactly the same holds for the Bose-Einstein condensation; see Ref.~\cite{Hanada:2020uvt} for details. 

For BEC, the off-diagonal long-range order (ODLRO)~\cite{PhysRev.104.576,RevModPhys.34.694} is often used to detect the genuine S$_N$ invariance~\cite{feynman_superfluidity1,feynman_superfluidity2,feynman_superfluidity3}. 
Let $\hat{\rho}=|\Phi\rangle\langle\Phi|$ be the $N$-particle density matrix made of the state $|\Phi\rangle\in{\cal H}_{\rm inv}$. From $\hat{\rho}$, the one-particle density matrix $\hat{\rho}_1$ is defined by tracing out $N-1$ particles as $\hat{\rho}_1={\rm Tr}_{2,\cdots,N}\hat{\rho}$.
We perform the spectral decomposition of $\hat{\rho}_1$ as 
\begin{align}
\hat{\rho}_1
=
n_{\rm max}|\Psi\rangle\langle\Psi|
+
\sum_i n_i|\Psi_i\rangle\langle\Psi_i|, 
\end{align}  
where $n_{\rm max}$ is the largest eigenvalue of $\hat{\rho}_1$. 
If $n_{\rm max}$ is of order one, $\langle x|\hat{\rho}_1|y\rangle$ does not vanish at long distance, and the system has the ODLRO. 
This $n_{\rm max}$ counts the number of degrees of freedom in BEC. For example, for the ground state 
$|\Phi\rangle=|\vec{0},\cdots,\vec{0}\rangle$, we obtain $\hat{\rho}_1=|\vec{0}\rangle\langle\vec{0}|$, hence $n_{\rm max}=1$, because
\begin{align}
\hat{\rho}_1={\rm Tr}_{2,\cdots,N}(|\vec{0},\cdots,\vec{0}\rangle\langle\vec{0},\cdots,\vec{0}|)=|\vec{0}\rangle\langle\vec{0}|.
\end{align}
On the other hand, if all $\vec{n}_i$'s are different, $n_{\rm max}=\frac{1}{N}$, because
 \begin{align}
\hat{\rho}_1
=
{\rm Tr}_{2,\cdots,N}\left(\frac{1}{N!}\sum_{\sigma,\sigma'}
|\vec{n}_{\sigma(1)},\cdots,\vec{n}_{\sigma(N)}\rangle\langle\vec{n}_{\sigma'(1)},\cdots,\vec{n}_{\sigma'(N)}|
\right)
=
\frac{1}{N}\sum_{i=1}^N
|\vec{n}_i\rangle\langle\vec{n}_i|.
\end{align}

This $n_{\rm max}$ is related to the Polyakov loop as follows~\cite{Hanada:2020uvt}. Firstly note that the group element $\sigma$ in \eqref{BEC-partition-function} is the Polyakov loop operator. In the thermodynamic limit, such an element $\sigma\in{\rm S}_N$ that leaves a typical state dominating the partition function invariant gives the expectation value. If $N_0=N-M$ particles are in the Bose-Einstein condensate, a typical state has a S$_{N_0}$-permutation symmetry.  
Any element of S$_{N_0}$ is a product of cyclic permutations. The eigenvalues of a cyclic permutation of length $k$ is $e^{2\pi i l/k}, l=1,2,\cdots,k$. 
As $N_0\to\infty$, dominant cyclic permutations becomes infinitely long~\cite{feynman_superfluidity1,feynman_superfluidity2,feynman_superfluidity3}, and the constant offset of the distribution of Polyakov loop phases proportional to $N_0$ appears. 
When $N_0=N\to\infty$, the phase distribution becomes completely uniform. 
See Ref.~\cite{Hanada:2020uvt} for more details. 

\subsection{Speculations regarding the Maldacena-Milekhin conjecture}\label{sec:gauged-vs-ungauged}
\hspace{0.51cm}
In Ref.~\cite{Maldacena:2018vsr}, Maldacena and Milekhin conjectured that the gauge-singlet constraint is not important at the low-energy regime of the D0-brane matrix model.
That is, the `gauged' partition function we have been discussing,  
\begin{align}
Z_{\rm gauged}(T)
=
{\rm Tr}_{{\cal H}_{\rm inv}}(e^{-\hat{H}/T})
=
\frac{1}{{\rm volU}(N)}\int dU{\rm Tr}_{{\cal H}_{\rm ext}}(\hat{U}e^{-\hat{H}/T}), 
\end{align}
should be exponentially close to the `ungauged' partition function 
\begin{align}
Z_{\rm ungauged}(T)
=
{\rm Tr}_{{\cal H}_{\rm ext}}(e^{-\hat{H}/T})
=
\frac{1}{{\rm volU}(N)}\int dU{\rm Tr}_{{\cal H}_{\rm ext}}(e^{-\hat{H}/T})  
\end{align}
at large-$N$.\footnote{
A similar characterization is valid for the microcanonical partition function, which can be applied to the M-theory parameter region where the gravity dual is the eleven-dimensional Schwarzschild black hole.
}
Specifically, the difference of the free energy should decay as $\sim\exp(-C\lambda^{1/3}/T)$, where $C$ is of order 1.
Therefore, the gauged and ungauged theory should be almost indistinguishable at $T\ll\lambda^{1/3}$, where weakly-coupled string or M-theory is a legitimate dual description. (Actually this conjecture was developed based on the intuition in the gravity side.) 
Monte Carlo simulation provided a result consistent with this conjecture~\cite{Berkowitz:2018qhn}. 

A natural mechanism in the matrix model side that can lead to this relation is that each low-energy state in ${\cal H}_{\rm ext}$ is invariant under a large subgroup of U$(N)$.\footnote{
More naive answer would be that only the U(1) subgroup, which acts on all the states trivially, leaves low-energy states invariant. 
However this possibility is excluded due to the mismatch of the distribution of the phases of the Polyakov loop: it gives the delta-function-like distribution, which is rather different from actual distribution. 
}
In usual confining gauge theory that has a mass gap of order $N^0$, this happens trivially at the energy scale well below the gap, because the ground state dominates the partition function.
A highly nontrivial point in the Maldacena-Milekhin conjecture is that they claim it happens even though the D0-brane matrix model does not have such gap; namely the non-singlet sector should be gapped while the singlet sector is not gapped. 
But perhaps we should not find it too surprising, because the same light mode, represented by a small block, can be excited multiple times. If the multiplicities of light modes are $n_1, n_2,\cdots$, then such a state in ${\cal H}_{\rm ext}$ is invariant under U($n_1$)$\times$U($n_2$)$\times\cdots$. If the multiplicities grow sufficiently fast as the energy goes down, it would be hard to distinguish the gauged and ungauged theories.

\bibliographystyle{utphys}
\bibliography{probe}

\end{document}